# Cyber Security Incident Handling, Warning and Response System for the European Critical Information Infrastructures (CyberSANE)

Spyridon Papastergiou[1], Haralambos Mouratidis[2] and Eleni-Maria Kalogeraki[1]

[1] Department of Informatics, University of Piraeus, 80 Karaoli and Dimitriou Str., 18534 Piraeus, Greece
paps@unipi.gr, elmaklg@unipi.gr

[2] School of Computing, Engineering and Mathematics, University of Brighton, Brighton BN2 4GJ, UK
H.Mouratidis@brighton.ac.uk

**Abstract.** This paper aims to enhance the security and resilience of Critical Information Infrastructures (CIIs) by providing a dynamic collaborative, warning and response system (CyberSANE system) supporting and guiding security officers and operators (e.g. Incident Response professionals) to recognize, identify, dynamically analyse, forecast, treat and respond to their threats and risks and handle their daily cyber incidents. The proposed solution provides a first of a kind approach for handling cyber security incidents in the digital environments with highly interconnected, complex and diverse nature.

## 1  Introduction

In the digital era, Critical Infrastructures (CIs) are operating under the premise of robust and reliable ICT components, complex ICT infrastructures and emerging technologies (e.g. IoT, Cloud Computing) and are transforming into Critical Information Infrastructures (CIIs) that can offer a high degree of flexibility and efficiency in the communication and coordination of advanced services. The increased usage of information technology in modern CIIs means that they are becoming more vulnerable to the activities of hackers and other perpetrators of cyber-related crime.

Over the last few years, it is a common phenomenon to see daily headlines describing major cyber-attacks or some new strain of malware or insidious social engineering technique being used to attack ICT infrastructures. In particular, CIIs have become lately targets for cyberattacks attracting the attention of security researchers, cyber-criminals, hacktivists (e.g. Anonymous, LulzSec) and other such role-players (e.g. cyber-spies). These cyber actors have significantly evolved their tactics, techniques and procedures to include next-generation malware toolkits available in various locations on the internet (e.g. deep web, dark web) and new data exfiltration methods that give them an asymmetric quantum leap in capability. In the past years, there have been a number of cybersecurity meltdowns and high-profile breaches affecting critical infrastructures, such as the recent ransomware attacks, WannaCry and WanaCrypt0r 2.0, which affected more than 230,000 computers in over 150 countries, In most cases, the adversaries targeted the organizations' interconnected infrastructures as a means of spreading their harmful malware to a broaderaudience. Obviously, the impact of a compromised CII can extend far beyond the corporate boundaries, putting not just individual organizations but also their dependent entities at risk.

In 2016, the Commission introduced the E.U. Directive NIS 2016 that enforces all CIIs to report to an appropriate Computer Security Incident Response Team (CSIRT) any incident having a substantial impact on the provision of their services. Unfortunately, these efforts mostly focused on providing just the legal basis and creating an assurance framework for boosting the cyber security culture across sectors which are vital for the EU economy and society and moreover rely heavily on ICTs. Nevertheless, there has been a lack of innovation to capture and correlate events and information associated with cyber-attacks in CIIs as well as lack of appropriate approaches that support and facilitate effective cooperation among the CIIs entities in terms of exchanging specific cybersecurity risk and threats information.

The paper proposes a state of the art solution, the CyberSANE system, which aims to improve the detection and analysis of cyber-attacks and threats on CIIs and increases the knowledge of the current cyber threat landscape. In particular, the CyberSANE system helps the organizations to raise their preparedness, improve their cooperation with each other, and adopt appropriate steps to manage security risks, report and handle security incidents. The rest of the paper is structured as follows. Section 2 and 3 present the related work and the main aspects of the proposed incident handling approach respectively. Section 4 describes the CyberSANE system and its key components. Section 5 illustrates the components data flows and the overall system operation; and finally Section 6 draws the conclusions.

## 2  Current Efforts of Incident Handling in Critical Information Infrastructures

The main goal of the security incident handling and response process is to define the main aspects and principles for coordinating the effort that should be applied in managing a security breach/incident/event [1,2]. In principle, choosing the right approach for incident handling proves to be complicated. In recent years, a number of security incident response approaches and frameworks [3,4,5,6,7,8,9,10,11,12,13,14] have been introduced by the research and industrial communities as well various standardization bodies. Although, many of these approaches provide specific technical guidelines, aiming to enhance the security incident response capabilities of the organizations, they present significant limitations. In particular, Grimes (2007) argues that most of the existing incident response approaches follow a linear process that is outdated and does not support the highly efficient capability that is required to handle and manage today's incidents. Therefore, a progression flaw exists in these processes, since if one phase in the linear process is not completed, the entire process cycle may stop midstream. [15] notes that current incident response processes are too focused on the containment, eradication, and recovery-related activities and usually ignore, skip or do not emphasize on other important steps of incident management, such as investigations actions. [16] proposal gives emphasis on proactive preparation and reactive learning to encourage security incident learning. [17,18,19] argue that the existing incident handling approaches do not provide adequate guidance on how to conduct effective forensic investigations. Hence, current methods' limitation to assist and guide the investigators in forensic evidence analysis, undermines the value of the evidence and fails to promote incident resolution.

In addition, the available security information and event management solutions lack significant reactive and post-incident capabilities for managing incidents and events in the scope of the ICT-based CIIs providing inadequate technical guidance to the incident response professionals on how to detect, investigate and reproduce attacks. As such, and despite the socioeconomic importance of tools and techniques for handling incidents there is still no easy, structured, standardized and trusted way to manage and forecast interrelated, cybersecurity incidents in a way that takes into account the heterogeneity and complexity of the CIIs and the increasingly sophisticated types of attacks. Therefore, there is a pressing need for devising novel systems for efficient CIIs incident handling and support thorough and common understanding of cyber-attack situations in a timely manner.

In a nutshell, the main limitations [20,21,22] of the existing approaches are the following: (i) the traditional linear incident response models are too slow, ineffective and do not support the highly efficient capability that is required to handle and manage today's incidents; (ii) focus mostly on the proactive element (i.e. provide assistance and information to help prepare, protect, and secure) of the incident management; (iii) current approaches do not provide enough insight into the underlying causes of the incident; (iv) poor provisions for incident planning; (v) undermine the value of forensic evidence possibly required for subsequent legal action; (vi) do not take into account the risk-related results produced by existing risk assessment methodologies.

## 3   CyberSANE Incident Handling Approach

The proposed incident handling approach aims to address the aforementioned limitations of these existing methodologies and tools, providing a step-by-step guidance to manage incidents and breaches on CIIs occurred due to cyber attacks. On this account, CyberSANE pursues to combine active approaches that are used to detect and analyse anomaly activities and attacks in real-time with reactive approaches that deals with the analysis of the underlying infrastructure to assess an incident in order to provide a more holistic and integrated approach to incident handling. In this vein, CyberSANE aims to enhance the incident detection capabilities of the existing methods described in the previous section with a more efficient, elastic and scalable reasoning approach. The main characteristics of the proposed approach are the following: (i) learning from unstructured data without the need to understand the content; (ii) identification of unusual activities that match the structural patterns of possible intrusions (instead of predefined rules); and (iii) automatic identification and adaption to a change of the underlying infrastructure.

CyberSANE treats the handling of a cyber incident as a dynamic experimental environment that can be optimized involving all relevant CIIs' operators and security experts. CyberSANE's approach is based on simulations to facilitate the evaluation and analysis of an identified incident and support the investigation decision making process in a rigorous manner. The pursuit of CyberSANE is to support incident handling process with advanced correlation capabilities in terms of accuracy and efficiency strengthening the rational analysis. In this context, CyberSANE relies on pioneering mathematical models (e.g. machine learning, deep learning and Global Artificial Intelligence (AI) techniques) for analysing, compiling, combining and correlating all incident-related information and data from different levels and contexts (e.g. taking into consideration information, data and opinions collected and analysed from existing risk assessment frameworks (including the CYSM, MEDUSA, MITIGATE and SAURON approaches [23,24,25], knowledge and information acquired from previous incident investigations as well as evidential data extracted from the compromised cyber systems). Thus, these techniques will be able to identify, extract and analyse the most relevant parts of the information related to the initial incident in order to find the relationships between the compromised devices/systems and these evidences.

Moreover, efficient simulation experiments for generating multi-order evidence dependencies have been used to generate and construct secure, reliable and valid chains of evidence anticipating how the attack is progressing. Additionally, taking into account the effects of a security incident, real-time insights, alerts and warnings will be produced to increase situation awareness, inform the CIIs' stakeholders about the effects of the events and guide them how to react.

The main contribution of the current approach that differentiates it from the aforementioned related incident handling proposals is that it combines existing machine learning techniques, such as clustering and hidden Markov models, with deep learning and Global Artificial Intelligence (AI) to develop an innovative way that optimizes the automatic analysis of huge amounts of events, information and evidence. To identify malicious actions in the cyber assets, such as abnormal behaviours, it combines both structured data (e.g. logs and network traffic) and unstructured data (e.g. data coming from social networks and dark web) in a privacy-aware manner. Furthermore, it adopts deep semantic analysis techniques together with Natural Language Processing (NLP) methods (e.g. Named Entity Recognition and Word Sense Disambiguation) to extract important information from multilingual security-related contexts, facilitating multilingual data generation and exploitation within the networked ecosystem.

## 4   CyberSANE Incident Handling system

CyberSANE is an advanced, configurable and adaptable, Security and Privacy Incident Handling system (CyberSANE system), towards effective security incident detection and handling. The main goal of the proposed system is to improve, intensify and coordinate the overall security efforts for the effective and efficient identification; investigation, mitigation and reporting of realistic multi-dimensional attacks within the interconnected web of cyber assets in the CIIs and security events. The proposed approach takes into consideration and addresses both technical and cognitive challenges.

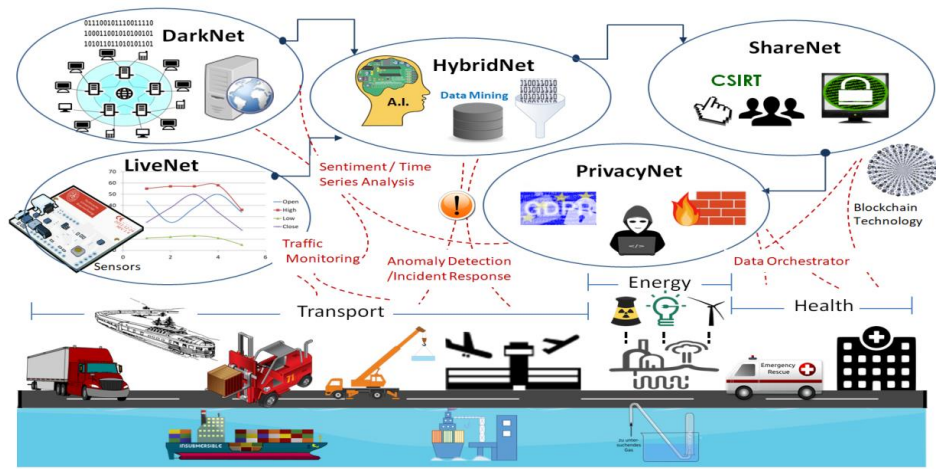

**Fig. 1.** CyberSANE Incident Handling system

From technical perspective, the system aims at collecting, compiling, processing and fusing all individual incident-related information ensuring their integrity and validity following the generic phases of ISO/IEC 27035:2016. In contrast, from cognitive point of view, the decision makers should be able to understand the technical aspects of an attack and draw conclusions on how to respond. In order to realize this vision, the CyberSANE system will be composed of five main components:

- The *Live Security Monitoring and Analysis (LiveNet)* component which is able to monitor, analyze, and visualize organizations' internal live network traffic in real time. This environment aggregates and visualizes traffic flowing through live networks as well as alerts given out by security appliances installed at critical points in the network with an overemphasis on threat prevention solutions.
- The *Deep and Dark Web mining and intelligence (DarkNet)* component which monitors the Dark and Deep Web in order to grasp and analyse the big picture of global malware/ cybersecurity activities.
- The *Data Fusion, Risk Evaluation and Event Management (HybridNet)* component that receives security related information on potential cyber threats from both LiveNet and Darknet respectively in order to analyze and evaluate the security situation inside an organization.

- The *Intelligence and Information Sharing and Dissemination (ShareNet)* component disseminates and shares information of useful incident-related information with relevant parties (e.g. industry cooperation groups, Computer Security Incident Response Teams - CSIRTs) about the effects and danger of incidents characterized diffusing threats.
- The *Privacy & Data Protection (PrivacyNet)* Orchestrator which provides a set of privacy (anonymization, pseudonymization, obfuscation), data protection orchestration and consistency capabilities.

It should be noted that the proposed solution and the incorporating techniques is able to operate in heterogeneous, large-scale, cross-border CIIs that are characterized by the following features: (i) complex, highly distributed, and large-scale cyber systems (including IoT and cyber-physical) with respect to the number of entities involved; (ii) heterogeneity of the underlying networks interconnecting the physical-cyber systems; and (iii) different levels of exposure to attacks. The following Sections provide a detailed description of the each component.

## 4.1 Live Security Monitoring and Analysis (LiveNet) component

The LiveNet is an advanced and scalable Live Security Monitoring and Analysis component capable of preventing and detecting threats and, in case of a declared attack, capable of mitigating the effects of an infection/intrusion. The main objective of this component is to implement the Identification, Extraction, Transformation, and Load process for collecting and preparing all the relevant information, serving as the interface between the underlying CIIs and the CyberSANE system. It includes proper cyber security monitoring sensors with network-based Intrusion Detection Systems (IDS), innovative Anomaly detection modules and endpoint protection solutions for accessing and extracting information, on a real-time basis, in order to detect complex and large-scale attacks (e.g. Advanced Persistent Threats). The incident-related information that reside in different and heterogeneous cyber systems may include various types of data, such as: active (unpatched) vulnerabilities in the technological infrastructure; misuse detection in the network or in the systems, including both host-based and network-based IDS deployment and integration; anomaly detection in the network or in the systems; system availability signals; network usage and bandwidth monitoring; industry proprietary protocol anomalies; SCADA vulnerabilities, etc.

LiveNet incorporates appropriate data management and reasoning capabilities for: near real-time identification of anomalies, threats, risks and faults and the appropriate reactions; (ii) proactive reaction to threats and attacks; and (iii) dynamic decision making in micro, macro and global level according to the end user's needs and the identified incidents/threats. These capabilities are empowered with more innovative algorithms based on techniques such as machine learning, deep learning and AI that identify previously unknown attacks. This component provides an abstraction of the collected information to the Data Fusion, Risk Evaluation and Event Management (HybridNet) component of the CyberSANE system. Moreover, all incidents-related information captured from LiveNet will be parsed, filtered, harmonized and enriched to ensure that only the data necessary for the multivariate and multidimensional analysis are available to the other components (e.g. HybridNet). Thus, LiveNet contributes as follows: (i) preventing a flood of irrelevant or repeated information from cluttering the HybridNet processing component; and (ii) consolidating the different data contents and formats towards a uniform perspective in order to provide the upper components a unified and convenient way to handle the information.

**4.2 Deep and Dark Web mining and intelligence (DarkNet) component**

The Deep and Dark Web mining and intelligence (DarkNet) component provides the appropriate Social Information Mining capabilities that will allow the exploitation and analysis of security, risks and threats related information embedded in user-generated content (UGC). This is achieved via the analysis of both the textual and meta-data content available from such streams. Textual information is processed to extract data from otherwise disparate and distributed sources that may offer unique insights on possible cyber threats. Examples include the identification of situations that can become a threat for the CIIs with significant legal, regulatory and technical considerations. Such situations are: organization of hacktivist activities in underground forums or IRC channels; external situations that can become a potential threat to the CIIs (e.g. relevant geopolitical changes); disclosure of zero day vulnerabilities; sockpuppets impersonating real profiles in social networks etc. Entities (e.g., events, places) and security-realated information will be uniquely extracted from textual content using advanced Natural Language Processing (NLP) techniques, such as sentiment analysis.

**4.3 Data Fusion, Risk Evaluation and Event Management (HybridNet) component**

The Data Fusion, Risk Evaluation and Event Management (HybridNet) component provides the intelligence needed to perform effective and efficient analysis of a security event based on: (i) information derived and acquired by the LiveNet and DarkNet components; and (ii) information and data produced and extracted from this component. In particular, HybridNet component retrieves incidents-related data via the LiveNet component from the underlying CIIs and data from unstructured and structured sources (e.g. from Deep and Dark Web) consolidated in a unified longitudinal view which are linked, analyzed and correlated ,in order to achieve semantic meaning and provide a more comprehensive and detailed view of the incident. In CyberSANE, a formal and uniform representation of digital evidence along with their relationships has been used to encapsulate all concepts of the forensic field and provide a common understanding of the structure of all information linking to evidence among the CIIs' operators and the forensics investigators. The main goal of the analysis process is to continuously carry out the assessment (e.g. identification of on-going attacks and related information, such as what is the stage of the attack and where is the attacker) and prediction (i.e. identification of possible scenarios of future attacks through forecasting models). HybridNet incorporates fusion models based on existing mathematical models (e.g. data mining, AI, deep learning, machine learning and visualization techniques). These models will support and provide reasoning capabilities for the near real-time identification of anomalies, threats and attacks, assessing any possible malicious actions in the cyber assets such as abnormal behaviors or malicious connections to identify unusual activities that match the structural patterns of possible intrusions. Once an attack is detected or predicted a simulation will be performed to form the full representation of the attack. In particular, a Security Incident/Attack Simulation Environment undertakes to generate and construct all secure, reliable and valid chains of evidence allowing: i) the identification of the attacker's behavior so far; ii) the identification of the attacker's goals and strategies and prediction of their next actions; and (iii) the accurate assessment of the impact of an incident on the CII and the damage caused so far.

The Security Incident/Attack Simulation Environment of the CyberSANE system comprises a set of novel mathematical instruments, including mathematical models for simulating, analyzing, optimizing, validating, monitoring simulation data and optimizing security incident handling process. Specifically, these instruments include: (i) a buddle of novel process/attack analysis and simulation techniques for designing, executing, analyzing and optimizing threat and attack simulation experiments that will produce appropriate evidence and information that facilitate the identification, assessment and mitigation of the CII-related risks; (ii) graph theory to implement attack graph generation, to perform security incident analysis and to strengthen the prognosis of future malefactor steps; (iii) pioneering mathematical techniques for analyzing, compiling and combining information and evidences about security incidents and attacks/threats patterns and paths in order to find relationships between the recovered forensic artefacts and piecing the evidential data together to develop a set of useful chain of evidence (linked evidence) associated with a specific incident; (iv) innovative simulation techniques which will optimize the automatic analysis of diverse data; (v) innovative techniques in order to link optimization and simulation. In this context, this simulation environment is fed with information about an incident and proceeds to calculate and generate a number of possible attack graphs (routes of possible attacks) and graphs of linked evidence (chains of evidence) and also compute probabilities for a sequence of events on top of these graphs. The resulting probabilistic estimate for the compromised CIIs' assets will be used to identify, model and represent the course of an attack as it propagates across the CIIs. It should be noted the HybridNet component continuously updates the simulation engine with information collected and piece of information, thereby enabling both understanding which assets might have been compromised, as well as gain more accurate estimates on the likelihood that other assets might be compromised in the future.

### 4.4 Intelligence and Information Sharing and Dissemination (ShareNet) component

The ShareNet component provides the necessary threat intelligence and information sharing capabilities within the CIIs and with relevant parties (e.g. industry cooperation groups, CSIRTs). It is responsible for the instantiation of the adopted intelligence model; in particular, ShareNet undertakes the identification and dissemination of, the right and sanitized information that have to be shared in a usable format and in a timely manner. This environment produces and circulates notifications containing critical information, enhancing the perception of the current situation and improving the projection into the future. It should be noted that all potential evidence from the systems that are suspected to be part of the infrastructure being investigated are forensically captured, stored and exchanged in a way that their integrity is maintained using the security and data protection methods of the PrivacyNet Orchestrator.

To this end, ShareNet follows a trusted and distributed intelligence and incident sharing approach to facilitate and promote the collaboration and secure and privacy-aware information sharing of the CIIs' operators with relevant parties (e.g. industry cooperation groups, CSIRTs), in order to exchange risk incident-related information, through specific standards and/or formats (STIX), improving overall cyber risk understanding and reduction. Privacy preserving is another important issue considered at every phase of sharing, applying methods such as anonymization or pseudo anonymization and encryption techniques incorporated in and made available from PrivacyNet Orchestrator. This brings forward a mixture of several cryptographic techniques that holds certain security guarantees.

**4.5 Privacy & Data Protection (PrivacyNet) Orchestrator**

Through the specific "Privacy & Data Protection Orchestrator" (PrivacyNet), it is possible to coordinate the abovementioned components of the CyberSANE system in order to ensure desired-levels of data protection for sensitive incident-related information, enabling the possibility to apply such protection in all phases of cyber security incident handling flow. The main purpose the PrivacyNet is to manage and orchestrate the application of the innovative privacy mechanisms and maximize achievable levels of confidentiality and data protection towards compliance with the highly-demanding provisions in the GDPR in the context of protecting sensitive incident-related information within and outside CIIs. To this end, PrivacyNet sets up the security and data "protection configurations" allowing security experts and members of the incident response team to specify all the protection steps that have to be performed and the required conditions to execute them, which can be referred to GDPR-based rules (and to other guidance for its application by the European Data Protection Board, formerly Art. 29 Data Protection Working Party).

In addition, the orchestration approach of the CyberSANE allows applying the most appropriate security and data protection methods depending on the user's privacy requirements, which cover a wide range of techniques including anonymization, location privacy, obfuscation, pseudonymization, searchable encryption, multi-party computation and verifiable computation, in order to meet the highly demanding regulatory compliance obligations, for example in relation to accountability towards data protection supervisory authorities, for adequate management of informed consent etc. For this reason, novel techniques and processes for enhancing the secure distribution and storage of all forensic artifacts in order to protect them from unauthorized deletion, tampering revision and sharing (e.g. Attribute-based Encryption (ABE) and block-chain technologies) have be combined.

## 5  CyberSANE Components Data Flow and System Operation

The abovementioned CyberSANE components communicate to carry out 13 operational data flows as illustrated in Figure 2 and outlines below. Through, those data flows and the overall operation, the CyberSANE system assists all the phases of security assessment and reduction of cyber risks on relevant operational scenarios.

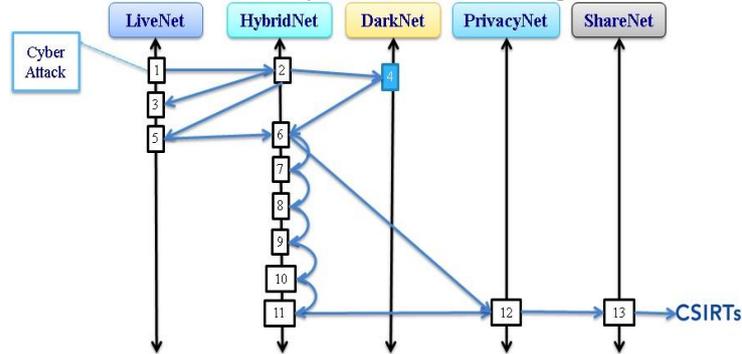

**Fig. 2.** CyberSANE System's & Components' Operation

The operational data flows supported by the CyberSANE system are the following: **(1)** Detected cyber attack visualized in the LiveNet Component; **(2)** Determine whether their organization has fallen victim to a cyber-attack; **(3)** Discovery, extraction and collection of raw data from various sources (e.g. servers, logs) and in different format **(4)** Extraction, harmonization and processing data from distributed sources (Dark Web, Social Media) that offer unique insights on the cyber threats and provide information about latest mechanisms of cyber-attacks; **(5)** Collected data is normalized, cleansed to remove redundant information and transformed into a common representation format; **(6)** All relevant information extracted are analyzed and correlated to provide a more comprehensive and detailed view of the incident; **(7)** Dependency evidence chains are generated; **(8)** Identification of on-going attacks (Identification of the attacker's behavior, prediction of next actions); **(9**) Evaluate the risks; Assess the impact and the assessment and cascading effects; and Formulate mitigation plan; **(10)** Prediction of possible scenarios of future attacks; **(11)** Visualization of incident related information enabling deep understanding of the situation and decision making (Notifications); **(12)** Secure and privacy aware managing and storing of incident-related information; and **(13)** Information sharing and dissemination of useful incident-related information.

## 6  Conclusions

The paper aims to leverage collected security information to find new ways of protection for technology assets, enabling the entity at risk to evaluate the risk and invest to limit that risk in an optimal way. Providing a way to securely collect both structured data (e.g. logs and network traffic) and unstructured data (e.g. data coming from social networks and dark web) and making them available for analysis fosters new innovations that will only unravel after having access to such data, harnessing its full potential. CyberSANE's has a twofold aim; to minimize the exposure to security risks/threats and help CIIs' operators to respond successfully to relevant incidents.

The ground-breaking nature of the proposed incident handling approach is based on: i) the identification of attacks and incidents using innovative approaches and algorithms of unobserved components techniques and linear state-space models producing meaningful information from cyber systems, ii) the combination of active incident handling approaches with reactive approaches producing real-time insights, alerts and warnings about cyber events, iii) innovative normalization process that unifies all relevant incident-related information gathered from heterogeneous CIIs, iv) novel attacks' scenarios and evidence representation with simulation techniques and visualization tools that increase the efficiency of investigation results, v) hybridization forms of mathematical models and combinations of data mining, Global artificial intelligence, machine learning that optimize evidential data from different sources.

The proposed CyberSANE System meets its objectives embedding core security features allowing faster and better operation of advanced cyber security functionalities. These aspects comprise an innovative, knowledge based, collaborative security and response dynamic system which increase the agility of the investigators and encourage continuous learning throughout the incident life cycle.

## Acknowledgements

The authors would like to thank the University of Piraeus Research Centre for its continuous support.

## References

1. West-Brown, M. J., Stikvoort, D., Kossakowski, K. P., Killcrece, G., & Ruefle, R. (2003a). Handbook for computer security incident response teams (csirts) (No. CMU/SEI-2003-HB-002). Carnegie-mellon univ pittsburgh pa software engineering inst.
2. Wiik, J., & Kossakowski, K. P. (2005). Dynamics of Incident Response. In 17th Annual FIRST Conference on Computer Security Incident Handling. Singapore.
3. British Standards Institution. (2011). Bs Iso/Iec 27035:2011 - Information Technology. Security Techniques. Information Security Incident Management.
4. Cichonski P, Scarfone K. (2012a) Computer security incident handling guide recommendations of the National Institute of Standards and Technology (NIST). Gaithersburg: NIST;
5. ENISA CSIRTs by Country-Interactive Map. Available at https://www.enisa.europa.eu/topics/csirts-in-europe/csirt-inventory/certs-by-country-interactive-map
6. Northcutt, S. (2003). Computer Security Incident Handling Version 2.3.1
7. Vangelos, M. (2011). "Incident Response: Managing," in Encyclopedia of Information Assurance. Taylor & Francis, pp 1442-1449.

8. Werlinger, R., Muldner, K., Hawkey, K., and Beznosov, K. (2010). "Preparation, Detection, and Analysis: The Diagnostic Work of It Security Incident Response," Information Management & Computer Security (18:1), pp. 26-42.
9. Khurana H, Basney J, Bakht M, Freemon M, Welch V, Butler R. Palantir. (2009). A framework for collaborative incident response and investigation. In: Proceedings of the 8th symposium on identity and trust on the internet. p. 38e51.
10. Grobauer B, Schreck T. (2010). Towards incident handling in the cloud. In: Proceedings of the 2010 ACM workshop on cloud computing security workshop (CCSW 10). p. 77-85.
11. Monfared, A., & Jaatun, M. G. (2012). Handling compromised components in an IaaS cloud installation. Journal of Cloud Computing:Advances, Systems and Applications 1,16.
12. Line, M. B. (2013). A case study: preparing for the smart grids-identifying current practice for information security incident management in the power industry. In IT Security Incident Management and IT Forensics (IMF), 2013 7 International Conference on IT security incident management and IT forensics (pp. 26-32). IEEE.
13. Cusick, J. J., & Ma, G. (2010). Creating an ITIL inspired Incident Management approach: Roots, response, and results. In Network Operations and Management Symposium Workshops (NOMS Wksps), 2010 IEEE/IFIP (pp. 142-148). IEEE.
14. Connell A, Palko T, Yasar H. Cerebro. (2013). A platform for collaborative incident response and investigation. In: 2013 IEEE international conference on technologies for homeland security (HST).
15. Ahmad, A., Hadgkiss, J., and Ruighaver, A.B. (2012). Incident Response Teams– Challenges in Supporting the Organisational Security Function. Computers & Security (31:5), pp. 643-652.
16. Shedden, P., Ahmad, A., and Ruighaver, A.B. (2011). Informal Learning in Security Incident Response Teams, in: 2011 Australasian Conference on Information Systems.
17. Casey, E. (2006). Investigating Sophisticated Security Breaches. Communications of the ACM (49:2), pp. 48-55.
18. Nnoli, H., Lindskog, D., Zavarsky, P., Aghili, S., and Ruhl, R. (2012). The Governance of Corporate Forensics Using Cobit, Nist and Increased Automated Forensic Approaches, in: 2012 International Conference on Privacy, Security, Risk and Trust. IEEE.
19. Tan, T., Ruighaver, T., and Ahmad, A. (2003). Incident Handling: Where the Need for Planning Is Often Not Recognised, in: 1st Australian Computer, Network & Information Forensics Conference.
20. FireEye. (2013). The Need for Speed: 2013 Incident Response Survey.
21. Grispos, G., Glisson, W. B., & Storer, T. (2014). Rethinking security incident response: The integration of agile principles. arXiv preprint arXiv:1408.2431.
22. Ab Rahman, N. H., & Choo, K. K. R. (2015). A survey of information security incident handling in the cloud. Computers & Security, 49, 45-69.
23. Papastergiou, S., & Polemi, D. (2017). Securing maritime logistics and supply chain: the medusa and mitigate approaches in proceedings of 2nd nmiotic conference on cyber security. Maritime Interdiction Operations Journal, 14(1), 42–48. ISSN: 2242-441X.
24. Papastergiou, S., & Polemi, N. (2018). MITIGATE: A dynamic supply chain cyber risk assessment methodology. In X. S. Yang, A. Nagar, & A. Joshi (eds), Smart trends in systems, security and sustainability. Lecture notes in networks and systems 18: 1–9. Springer.
25. Kalogeraki, E.-M., Papastergiou, S., Polemi N. "SAURON Real-life Use Cases: Terrorists attack a Cruise Ship berthed at a Port Facility" the 9th NMIOTC Annual Conference "Fostering Projection of Stability through Maritime Security: Achieving Enhanced Capabilities and Operational Effectiveness" June 5-7 2018.